\documentclass[%
 reprint,
superscriptaddress,
 amsmath,amssymb,
 aps,
]{revtex4-2}

\usepackage{graphicx}
\usepackage{dcolumn}
\usepackage{bm}
\usepackage[hyperindex,breaklinks]{hyperref}
\hypersetup{colorlinks=true}
\usepackage{breakurl}
\usepackage[mathlines]{lineno}
\usepackage[caption=false]{subfig}

\newcommand{\nuebar}{$\overline{\nu}_{e}$ }
\newcommand{\nuebars}{$\overline{\nu}_{e}$'s }
\newcommand{\alphan}{($\alpha, n$) }

\begin{document}

\preprint{APS/123-QED}


\title{Evidence of Antineutrinos from Distant Reactors using Pure Water at SNO+}

\author{ A.\,Allega}
\affiliation{\it Queen's University, Department of Physics, Engineering Physics \& Astronomy, Kingston, ON K7L 3N6, Canada}
\author{ M.\,R.\,Anderson}
\affiliation{\it Queen's University, Department of Physics, Engineering Physics \& Astronomy, Kingston, ON K7L 3N6, Canada}
\author{ S.\,Andringa}
\affiliation{\it Laborat\'{o}rio de Instrumenta\c{c}\~{a}o e  F\'{\i}sica Experimental de Part\'{\i}culas (LIP), Av. Prof. Gama Pinto, 2, 1649-003, Lisboa, Portugal}
\author{ J.\,Antunes}
\affiliation{\it Laborat\'{o}rio de Instrumenta\c{c}\~{a}o e  F\'{\i}sica Experimental de Part\'{\i}culas (LIP), Av. Prof. Gama Pinto, 2, 1649-003, Lisboa, Portugal}
\affiliation{\it Universidade de Lisboa, Instituto Superior T\'{e}cnico (IST), Departamento de F\'{\i}sica, Av. Rovisco Pais, 1049-001 Lisboa, Portugal}
\author{ M.\,Askins}
\affiliation{\it University of California, Berkeley, Department of Physics, CA 94720, Berkeley, USA}
\affiliation{\it Lawrence Berkeley National Laboratory, 1 Cyclotron Road, Berkeley, CA 94720-8153, USA}
\author{ D.\,J.\,Auty}
\affiliation{\it University of Alberta, Department of Physics, 4-181 CCIS,  Edmonton, AB T6G 2E1, Canada}

\author{ A.\,Bacon}
\affiliation{\it University of Pennsylvania, Department of Physics \& Astronomy, 209 South 33rd Street, Philadelphia, PA 19104-6396, USA}
\author{ N.\,Barros}
\affiliation{\it Laborat\'{o}rio de Instrumenta\c{c}\~{a}o e  F\'{\i}sica Experimental de Part\'{\i}culas (LIP), Av. Prof. Gama Pinto, 2, 1649-003, Lisboa, Portugal}
\affiliation{\it Universidade de Lisboa, Faculdade de Ci\^{e}ncias (FCUL), Departamento de F\'{\i}sica, Campo Grande, Edif\'{\i}cio C8, 1749-016 Lisboa, Portugal}
\author{ F.\,Bar\~{a}o}
\affiliation{\it Laborat\'{o}rio de Instrumenta\c{c}\~{a}o e  F\'{\i}sica Experimental de Part\'{\i}culas (LIP), Av. Prof. Gama Pinto, 2, 1649-003, Lisboa, Portugal}
\affiliation{\it Universidade de Lisboa, Instituto Superior T\'{e}cnico (IST), Departamento de F\'{\i}sica, Av. Rovisco Pais, 1049-001 Lisboa, Portugal}
\author{ R.\,Bayes}
\affiliation{\it Laurentian University, School of Natural Sciences, 935 Ramsey Lake Road, Sudbury, ON P3E 2C6, Canada}
% ...

\collaboration{The SNO+ Collaboration}

\begin{abstract}
\newpage
The SNO+ Collaboration reports the first evidence of reactor antineutrinos in a Cherenkov detector.  The nearest nuclear reactors are located 240~km away in Ontario, Canada.  This analysis uses events with energies lower than in any previous analysis with a large water Cherenkov detector.  Two analytical methods are used to distinguish reactor antineutrinos from background events in 190 days of data and yield consistent evidence for antineutrinos with a combined significance of 3.5$\sigma$.  
\end{abstract}

\maketitle

\textit{Introduction}.  
Measurements of reactor antineutrinos have made crucial contributions to 
the study of neutrinos, including the discovery of the neutrino~\cite{Cowan:1956rrn} and leading measurements of neutrino oscillation parameters $\theta_{13}$, $\Delta m^2_{32}$, and $\Delta m^2_{21}$~\cite{DayaBay:2018yms, RENO:2018dro, KamLAND:2013rgu}.  
These measurements used organic liquid scintillators to detect the products of inverse beta decays (IBDs) on protons: $\overline{\nu}_{e} + p \rightarrow e^+ + n$.  
The $e^+$ carries most of the energy of the \nuebar and produces a prompt signal while the $n$ is captured by a nucleus as it thermalizes, producing a delayed signal, typically of one or more $\gamma$'s depending on the capture isotope.  
Water Cherenkov detectors have used different neutrino interaction channels and also made invaluable contributions to the study of neutrinos, including the discovery of neutrino oscillations~\cite{Super-K:1998Atm,SNO:2002tuh} and leading measurements of parameters $\theta_{12}$, $\theta_{23}$, and $\Delta m^2_{32}$~\cite{SNO:2011hxd, Super-K:2016solar, IceCube:2017lak, Super-Kamiokande:2017yvm, T2K:2019bcf}.

Previous water Cherenkov detectors have not identified reactor \nuebars largely due to their detector thresholds, which have provided a low efficiency to detect the 2.2-MeV $\gamma$ emitted when a neutron is captured by a hydrogen nucleus.  
One approach to improving the identification of neutrons is to make use of a higher-energy capture signal, for example, by dissolving Cl or Gd into the water~\cite{SNO:2003bmh, Super-Kamiokande:2021acd, Super-Kamiokande:2021the}. 
Another approach is to attain lower thresholds, though this would also admit the more abundant radioactive backgrounds below 3~MeV~\cite{Super-Kamiokande:2008n}.  
Distinguishing signals and backgrounds below 3~MeV is challenging in water Cherenkov detectors due to their coarse energy resolution ($\approx$25\% at 3~MeV for SNO+), which arises from small numbers of detected photons.  

SNO+ has achieved the lowest energy threshold of any large Cherenkov detector, at approximately 1.4~MeV for an electron at the center of the detector, which yields an efficiency around 50\% to detect the 2.2-MeV $\gamma$~\cite{SNO:2020bdq}.  
The analysis presented in this Letter addresses the higher rate of radioactivity at these lower energies using two analytical methods.  
Both methods suppress the accidental background from ambient radioactivity by more than 4 orders of magnitude while maintaining a relatively high IBD efficiency. 
All relevant backgrounds, including \alphan reactions and atmospheric neutrino interactions, are estimated with data in ``sidebands'' that are complementary to the IBD region of interest. 

\textit{Detector \& data}.  
SNO+ is a multipurpose neutrino experiment located at SNOLAB, 2~km underground in an active mine, near Sudbury, Ontario.  The detector consists of a spherical acrylic vessel (AV) that is 6.0 meters in radius, submerged in ultrapure water, and surrounded by an array of 9,362 eight-inch photomultiplier tubes (PMTs) at a radius of 8.5~m.  From May 2017 to July 2019, the AV was filled with 905 tons of ultrapure water and from September 2017, SNO+ operated as a low-threshold Cherenkov detector.  The detector is described in Ref.~\cite{SNO:2021xpa}.  

Data from the SNO+ low-threshold water phase are divided into two sets: the first has higher rates of internal radioactivity due to radon ingress into the detector, while the second has significantly lower background rates owing to the installation of an N$_{2}$ cover gas system on the top of the detector in September 2018. This analysis used the lower-background dataset, which has a total live time of 190.3~days. The first dataset, which has 140.7~days of live time, was analyzed to provide additional statistics in background sideband studies.

\textit{Signal prediction}.  
The rate and energy spectrum of reactor IBDs were predicted using the Huber-Mueller isotope model and other inputs, as described in Ref.~\cite{DayaBay:2016ssb}.  
The flux of \nuebar originated from 18 reactor cores in Canada plus approximately 100 cores in the USA, with an average baseline around 620~km (baselines were weighted by the expected IBDs).  The three nearest reactor complexes (Bruce, Pickering, and Darlington at distances of 240, 340, and 350~km) house Canada Deuterium Uranium (CANDU) reactors and were estimated to have yielded nearly 60\% of the IBDs.  
The thermal powers of the CANDU reactors were modeled using hourly electrical power provided by IESO~\cite{IESO}. All other reactors were modeled using their respective thermal powers, provided annually as monthly averages by the IAEA~\cite{PRIS}. 

Changes in the relative fractions of fissile isotopes during fuel burn-up result in a time evolution of the emitted \nuebar energy spectrum and flux. 
However, because CANDU reactors are constantly refueled and \nuebar were produced by a large number of reactors, the resulting variation in total flux was $<$ 1\%.  
As such, constant fission fractions were used in the predictions for the CANDU pressurized heavy water reactors (PHWRs) as well as pressurized/boiling water reactors (P/BWRs):
($^{235}$U, $^{239}$Pu, $^{238}$U, $^{241}$Pu) were set to
(0.52, 0.42, 0.05, 0.01) for PHWRs~\cite{AECL:2013} and to 
(0.568, 0.297, 0.078, 0.057) for P/BWRs~\cite{KamLAND:2002uet}.

Known biases in the flux model were corrected by scaling the prediction by 0.945$\pm$0.007 to match the global average of flux measurements~\cite{DayaBay:2018heb}.  
Because the total systematic uncertainty on the prediction is negligible relative to the statistical uncertainties of the dataset, uncertainty components were largely taken from Ref.~\cite{DayaBay:2016ssb} and references therein, totaling to $\pm$3\% in the rate of IBDs.  

The survival probability of \nuebars that reached the SNO+ detector was dominantly determined by the neutrino mass splitting $\Delta m_{21}^{2}$ and oscillation angle $\theta_{12}$.  
Oscillation parameters from a recent global fit~\cite{ParticleDataGroup:2018ovx} were used and the net reduction in detected IBDs due to oscillation was found to be 46\%. 
The uncertainties in $\Delta m_{21}^{2}$ and $\sin^{2}\theta_{12}$ were propagated to the expected IBD rate, resulting in a $\pm$4\% uncertainty.  

The expected number of IBDs within the 2.5~ktons of water enclosed by the array of PMTs during the 190.3-day dataset was thus calculated to be 160.4$^{+8.1}_{-8.4}$.

\textit{Analysis methods}.  
Two analytical methods, likelihood ratio (LR) and boosted decision tree (BDT), were used to identify coincidences of an IBD positron (``prompt'' event) and neutron (``delayed'' event).  The two methods used a common set of criteria to remove instrumental backgrounds and electronic noise as described in Ref.~\cite{SNO:2018ydj}, and then applied slightly different initial IBD selection criteria as described below and summarized in Table~\ref{tab:cuts}.  

To avoid processing the large number of events from radioactivity that were below the energy threshold, events were reconstructed only if the number of PMT hits (Nhits) was greater than or equal to 15 (roughly 2~MeV). 
Additionally, to ensure that neutrons could be selected, any event that followed within 1~ms of an event with Nhits $\geq$ 15 was reconstructed (the neutron capture time constant is 207~$\mu$s~\cite{SNO:2020bdq}). 
The reconstruction of event position, direction, and energy, along with associated systematic uncertainties, is described in Ref.~\cite{SNO:2018ydj}.  
Events with poorly reconstructed energies or positions were removed using several figures of merit (FOMs).  
Two additional parameters were used to evaluate the compatibility of each event with a Cherenkov signal~\cite{SNO:2018ydj}.  
The first, in-time ratio (ITR), is the fraction of Nhits with time-of-flight-corrected hit times within [-2.5, 5]~ns, which identified events with broad time distributions or poorly reconstructed positions.  
The second, $\beta_{14}$, quantifies the spatial isotropy of the hit PMTs using the first and fourth Legendre polynomials of the distribution of angles between the PMTs with respect to the reconstructed position. Events with a spatially uniform distribution of hit PMTs would have values of $\beta_{14}$ near zero.  

The reconstructed position $\mathbf{r}$ of a prompt event was required to be in one of two fiducial volumes (FVs): inside the AV (internal) and between the AV and the PMTs (external).  The region close to the AV was excluded to avoid \alphan reactions from the acrylic.  
This analysis is the first from SNO+ to use the external volume, where the energy scale has been calibrated as a function of event position and direction, and reconstruction systematic uncertainties have been evaluated similarly to those for the volume inside the AV.  

Loose bounds were applied to the reconstructed energy $E$ of both prompt and delayed events.  
The distance between prompt and delayed events $\Delta r$ was also loosely bound. 
The BDT method instead used Nhits$_{p,12}$ for delayed events, which is the maximum number of PMT hits found in a sliding time window of 12~ns, and was calculated using the time-of-flight-corrected PMT hit times assuming the prompt event position~\cite{SNO:2020bdq}.  Thus, a cut on Nhits$_{p,12}$ effectively cut on both Nhits and $\Delta r$, without directly relying on a reconstructed position for the delayed event.  Delayed events with Nhits$_{p,12} >$ 6 were selected to suppress low-energy accidentals.  

The time between prompt and delayed events $\Delta t$ excluded the first 3~$\mu$s after each prompt event based on one of the instrumental background criteria aimed at avoiding sequential detector triggers due to high-energy events or electronics noise. 
Any prompt event with more than one delayed event within 500~$\mu s$ was rejected, which reduced backgrounds that have high neutron multiplicities, such as atmospheric neutrino interactions, while sacrificing a negligible fraction of IBDs. 
A 100-ms veto was applied after any event with Nhits $>$ 300 (roughly 40~MeV), which reduced backgrounds from charged-current atmospheric neutrino interactions. 
Cosmogenic muon products were avoided by vetoing all events within 20~s after identified muons.  Such a long veto window resulted in a small live-time sacrifice since only about three muons pass through the detector every hour.

\begin{table}[t]
    \centering
    \caption{Initial IBD selection criteria used by LR and BDT methods.  See text for details. ``NA'' means ``Not Applied''.}
    \label{tab:cuts}
    \begin{tabular}{lcccc} \hline \hline 
         & \multicolumn{2}{c}{LR} & \multicolumn{2}{c}{BDT} \\ \hline 
         & Prompt & Delayed & Prompt & Delayed \\ \hline
         Nhits & $\geq$ 15 & NA & $\geq$ 15 & $\leq$ 25 \\ 
         $\Delta t$ [$\mu$s] & \multicolumn{2}{c}{(3, 500)} & \multicolumn{2}{c}{(3, 500)} \\  
         ITR & $>$ 0.5 & NA & $>$ 0.5 & NA \\
         $\beta_{14}$ & (-0.6, 1.6) & NA & (-0.6, 1.6) & NA \\
         $||\mathbf{r}||$ (internal) [m] & $<$ 5.7 & $<$ 5.7 & $<$ 5.6 & NA \\
         $||\mathbf{r}||$ (external) [m] & (6.3, 7.5) & (6.1, 7.6) & (6.4, 7.3) & NA \\
         $z$ (external) [m] & \multicolumn{2}{c}{(-5.0, 5.0)} & \multicolumn{2}{c}{(-5.0, 5.0)} \\ 
         $E$ [MeV] & (2.5, 9.0) & $<$ 4.0 & (2.5, 9.5) & NA \\ 
         $\Delta r$ [m] & \multicolumn{2}{c}{$<$ 3.0} & \multicolumn{2}{c}{NA} \\ 
         Nhits$_{p,12}$ &  & NA &  & $>$ 6 \\ \hline \hline
    \end{tabular}
\end{table}

The neutron selection efficiency was measured using a deployed AmBe source~\cite{SNO:2020bdq} in order to correct for the imperfect modeling of the detector trigger around threshold. 
The source produces coincident pairs of a 4.4-MeV $\gamma$ and a neutron.  
Events were selected with high purity using the same criteria as in the IBD analysis. The resulting volume-weighted ratios of efficiencies (data/simulation) were applied as corrections to all predictions involving neutrons: 
0.85 $\pm$ 0.16 (LR) and 0.89 $\pm$ 0.23 (BDT).  
The large uncertainties arose from the variation of the ratios across the detector and the lack of complete deployment coverage across the FVs. These are the dominant uncertainties in the IBD prediction.  

The total AmBe-corrected efficiency of the initial IBD selection criteria was 4.9\% (LR) and 5.2\% (BDT).  For events that triggered the detector and satisfied the FV, $E$, and Nhits criteria, the initial IBD selection efficiency was about 60\% for both methods.
Individual selection efficiencies for the LR method are given in Table~\ref{tab:eff}. 

After applying all initial selections, the dominant background was found to be from accidental coincidences, at an estimated rate around 58 per day, or more than 10$^{4}$ accidentals in the 190-day dataset. Thus, it was necessary that cuts on the LR and BDT distributions reduce the accidental rate by at least 4 orders of magnitude without sacrificing a significant fraction of the expected IBDs.  
The number of IBDs and accidentals is tabulated for the LR method in Table~\ref{tab:eff}.  

\begin{figure*}[!t]

\subfloat[\label{subfig:likelihood}]{%
  \includegraphics[width=0.333\textwidth]{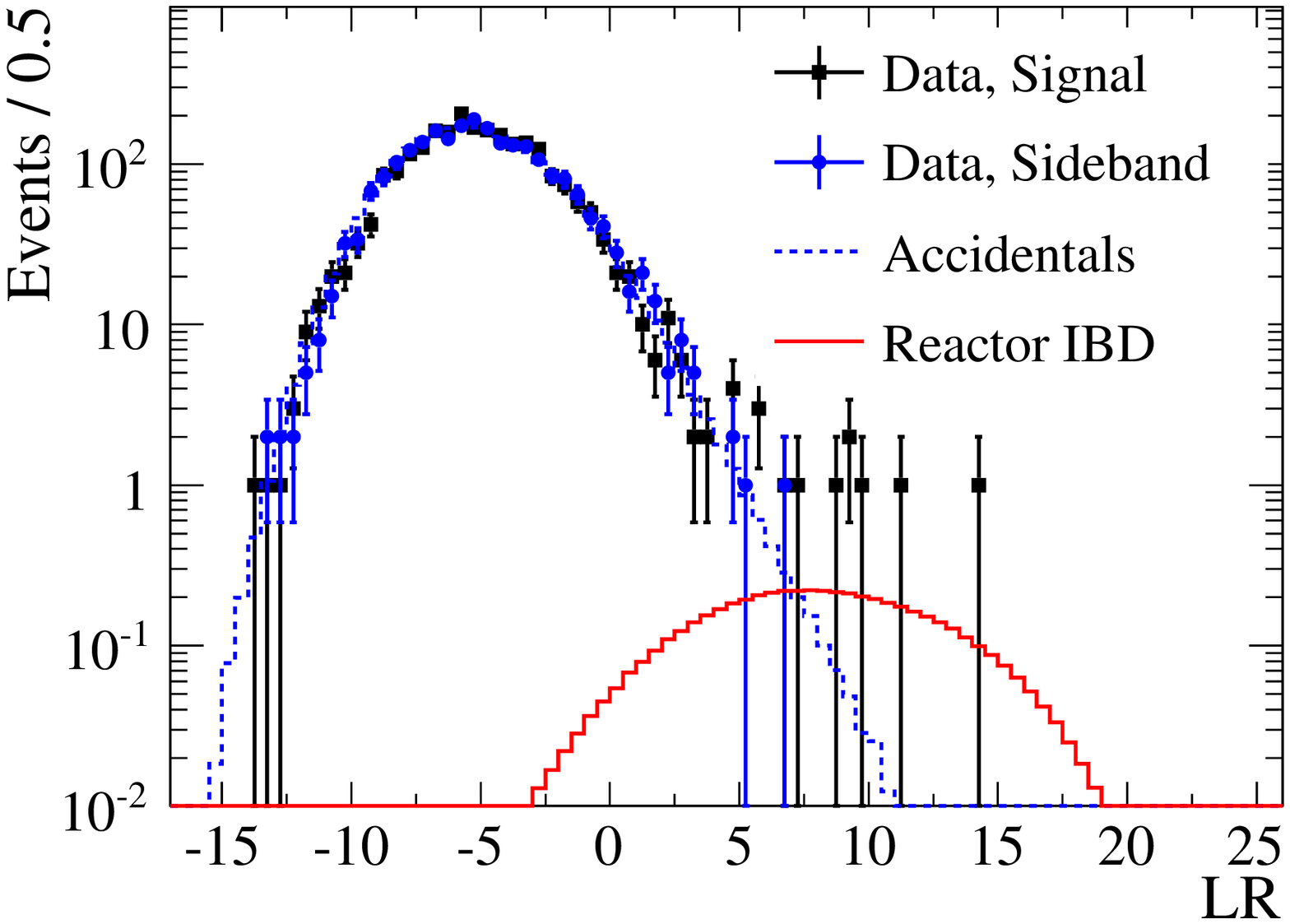}%
}\hfill
\subfloat[\label{subfig:likelihood_deltat}]{%
  \includegraphics[width=0.333\textwidth]{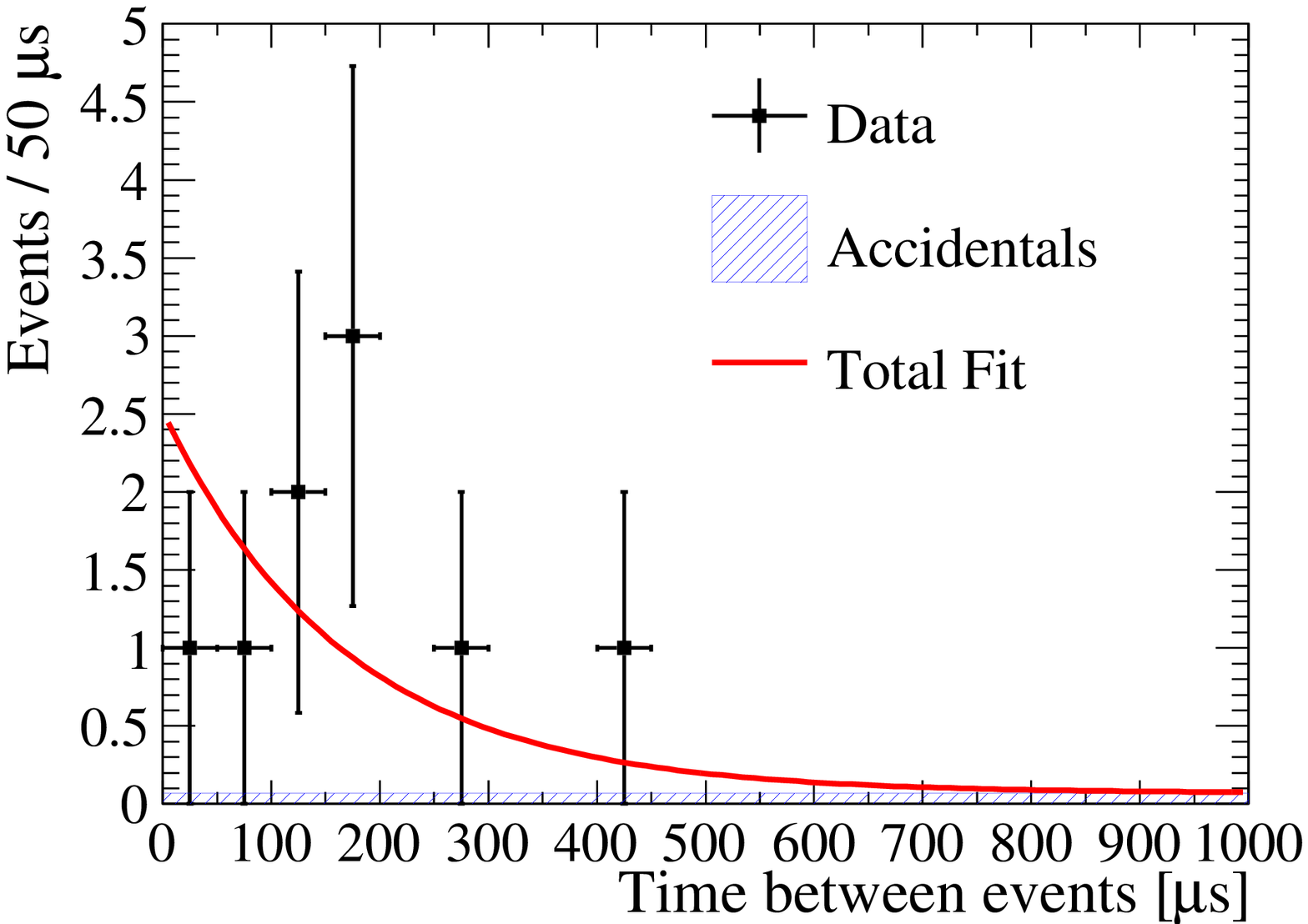}%
}\hfill
\subfloat[\label{subfig:likelihood_energy}]{%
  \includegraphics[width=0.333\textwidth]{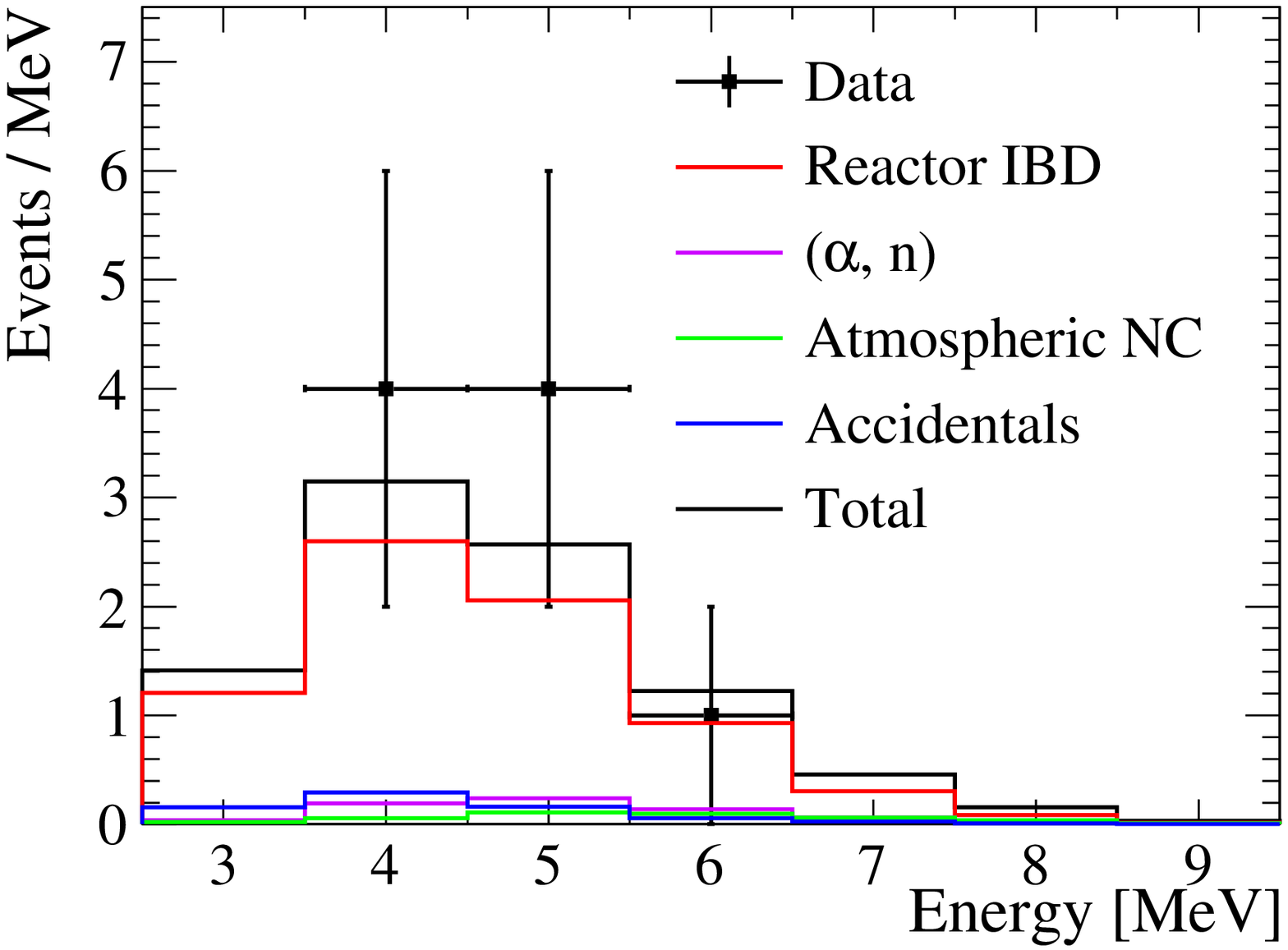}%
}

\subfloat[\label{subfig:bdt}]{%
  \includegraphics[width=0.333\textwidth]{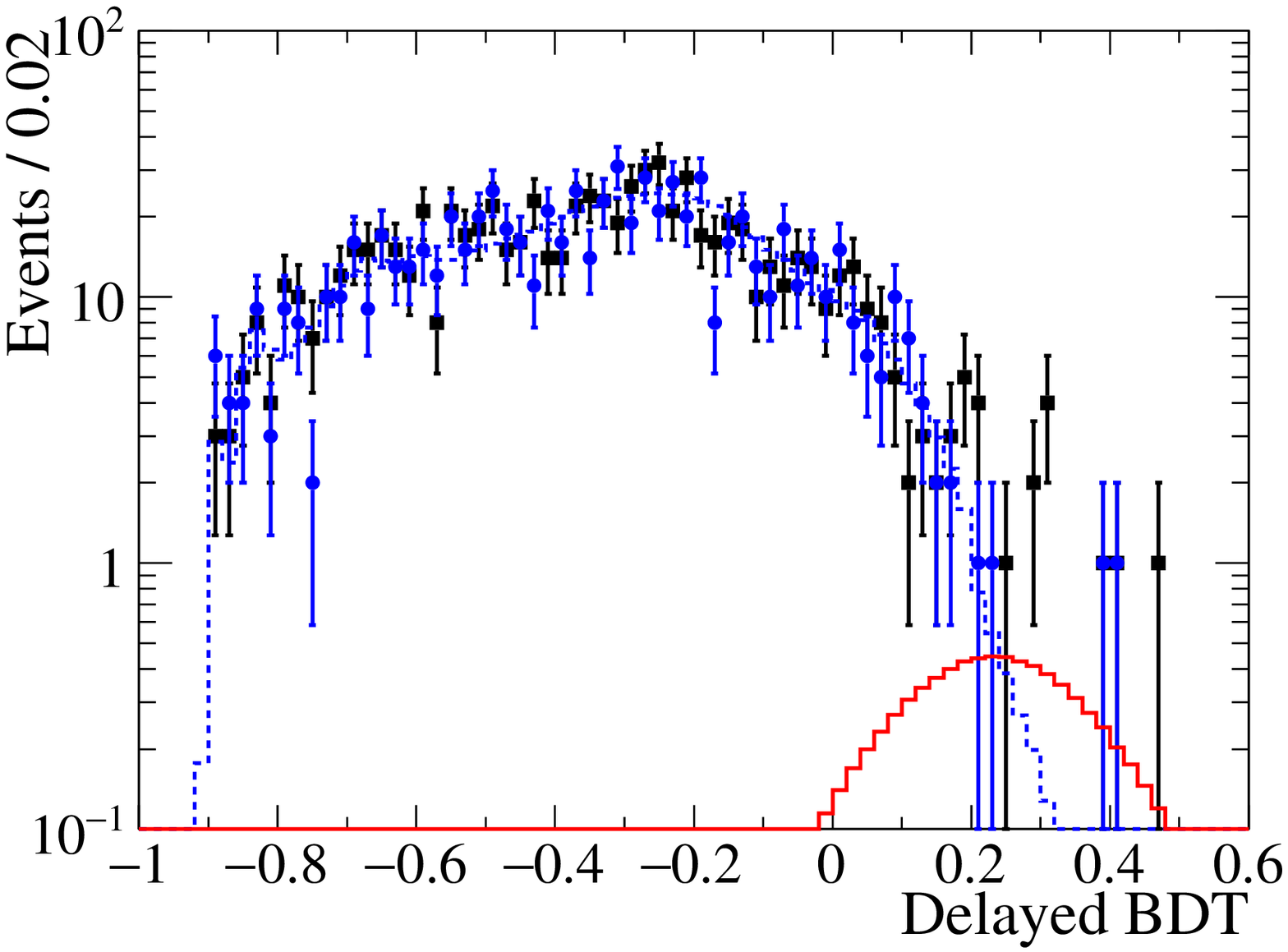}%
}\hfill
\subfloat[\label{subfig:bdt_deltat}]{%
  \includegraphics[width=0.333\textwidth]{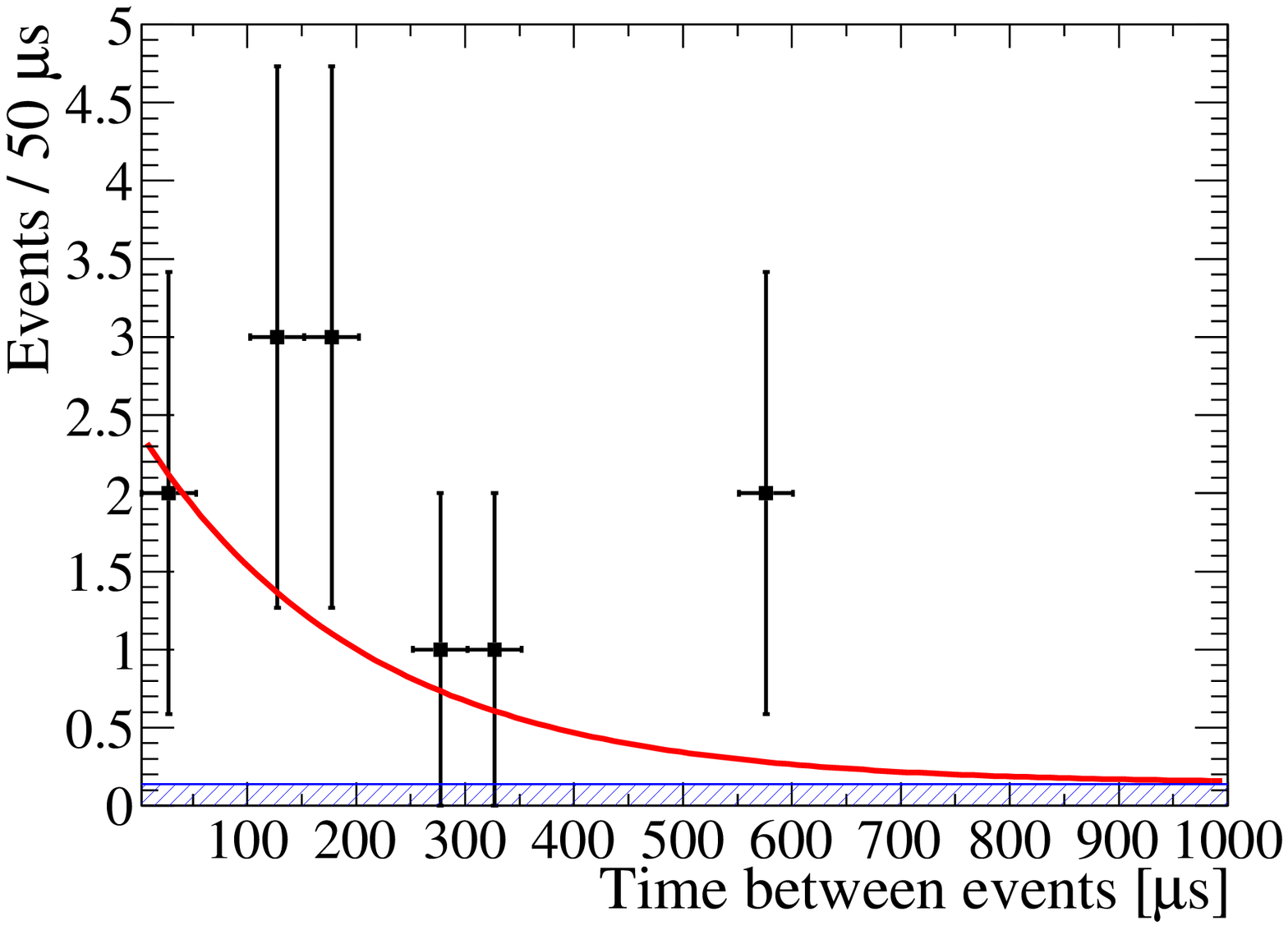}%
}\hfill
\subfloat[\label{subfig:bdt_energy}]{%
  \includegraphics[width=0.333\textwidth]{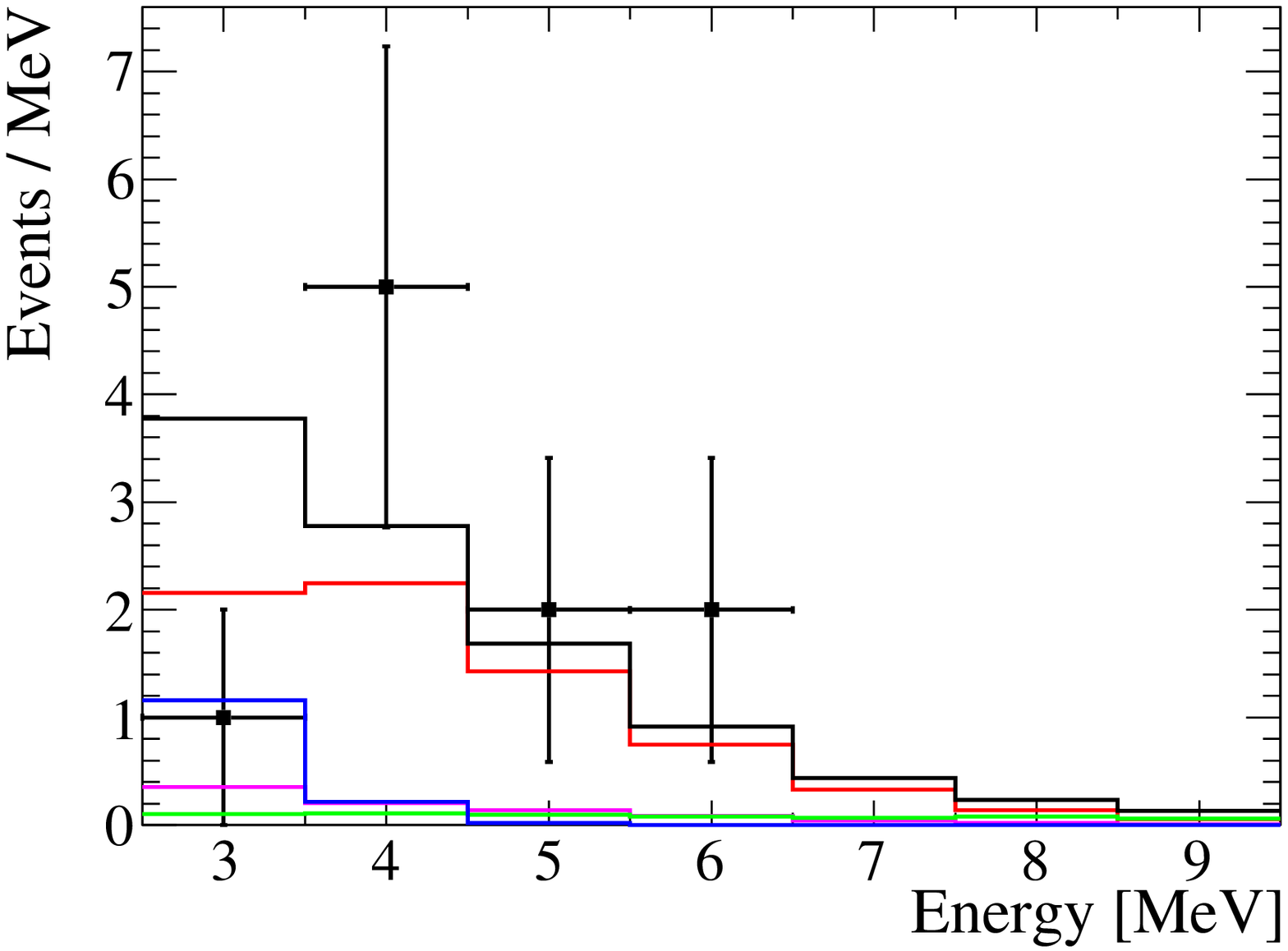}%
}

\caption{The top row shows the LR method and the bottom row shows the BDT method.  {\it Left:} The LR for internal events (a) and the delayed BDT for all events (d) that pass the criteria in Table~\ref{tab:cuts}. The signal (red) is normalized to the expectation. For both methods there is good agreement between the data in the $\Delta t$ sideband and the accidental estimation.  {\it Center:} The $\Delta t$ distribution for the observed events, including the region used as a sideband between 500 and 1000~$\mu$s. The fits include exponential and flat components to model the neutron captures and accidental backgrounds, respectively.  In the LR method (b), the accidental background is not expected to be flat because $\Delta t$ is included in the LR, but making this simplifying assumption for the fit has no significant impact~\cite{TK2020}.  {\it Right:} The prompt energy distributions for the selected events compared to the expectation for the signal and backgrounds.  The background components are normalized to their expected counts, while the IBD signal is normalized to the total number of observed events minus the total number of expected background events.} \label{fig:all_plots}
\end{figure*}

The LR method constructed a likelihood ratio using probability-density histograms for reactor IBDs and accidentals. The former were generated from reactor \nuebar simulations with SNO+ RAT (a {\sc Geant}4-based \cite{GEANT4:2002zbu} simulation package) and the latter were constructed by assigning a uniformly distributed random value for $\Delta t \in$ [0,1000]~$\mu$s to a pair of prompt- and delayed-like events, each randomly sampled from the data after applying their respective criteria in Table~\ref{tab:cuts}. 
Correlations between variables were accounted for by using two-dimensional histograms.  Event energy $E$ vs $\beta_{14}$ accounted for the energy dependence of $\beta_{14}$ that results from the broadening of the angular distribution of Compton-scattered $\gamma$'s with decreasing energy.  
Noting that the backgrounds have prompt $\gamma$'s, it also distinguished the broader angular distribution of $\gamma$'s from that of IBD $e^+$'s.  
A histogram of radial position $||\mathbf{r}||$ vs. radial direction cosine $\mathbf{u} \cdot \mathbf{r}$ (the unit dot product of event direction $\mathbf{u}$ and position $\mathbf{r}$) distinguished the isotropic and uniformly distributed IBDs from accidentals, which largely arose from $\gamma$'s emitted from the PMTs' glass, and were therefore relatively inward-pointing and at higher radii.  
Histograms for the $\Delta t$ and $\Delta r$ distributions were also included.  
Distinct sets of probability-density histograms were used for the internal and external volumes and a cut on each of the two resulting LR distributions was optimized separately.  Distributions of these variables are shown in Ref.~\cite{supp} and more details of this method are given in Ref.~\cite{TK2020}.  
The LR distribution for internal IBDs is shown as a red histogram in Fig.~\ref{fig:all_plots} (a).  

\begin{table}[t]
    \centering
    \caption{For the LR method, expected IBD efficiency of each initial selection, and IBD and accidental counts in 190 live days after each initial selection.  The last row shows these for the final selection based on the LR cuts.  Efficiencies are calculated for prompt and delayed events together. The volume-weighted correction from the AmBe source (0.85) is included in the trigger efficiency.}
    \label{tab:eff}
    \begin{tabular}{lccc} \hline \hline
    Initial selection & Efficiency [\%] & ~IBDs~ & Accidentals \\ \hline
    None &  & 160.4 &  \\  
    Trigger & 32.9 & 52.8 &  \\  
    Instrumentals~\cite{SNO:2018ydj} & 95.7 & 50.5 &  \\
    Nhits & 56.9 & 28.7 &  \\ 
    Valid reconstruction & 87.2 & 25.1 &  \\
    $\Delta t$ & 90.7 & 22.7 & 6.03 $\times$ 10$^6$ \\
    FOMs, ITR, $\beta_{14}$ & 88.8 & 20.2 & 1.86 $\times$ 10$^6$ \\
    Fiducial volume & 50.0 & 10.1 & 5.86 $\times$ 10$^5$ \\ 
    $E$ & 82.1 & 8.3 & 2.61 $\times$ 10$^5$ \\ 
    $\Delta r$ & 94.7 & 7.8 & 1.11 $\times$ 10$^4$ \\ \hline
    LR selection & 44.7 & 3.5 & 0.7 \\ \hline  \hline 
    \end{tabular}
\end{table}

The BDT method includes several variables in addition to those used in the LR method and naturally accounts for correlations between all variables. 
However, the most important difference with the LR method is that it uses distinct trees for prompt and delayed events in contrast to the single likelihood ratio used in the LR method.  This led to a significantly different selection of prompt-delayed pairs in the same dataset.  Reactor \nuebar simulations and accidental samples constructed from the data were used to train the BDTs.  The accidental samples were constructed similarly to those of the LR method except that $\Delta t$ was not a parameter in either BDT.  The prompt event BDT used five input variables: $E$, $\beta_{14}$, vertical position $z$, transverse position $\rho \equiv \sqrt{x^2+y^2}$, and radial direction cosine $\mathbf{u}\cdot \mathbf{r}$. 
A total of 12 variables were input to the delayed BDT. Similar to the LR method, the delayed BDT used $E$, $\rho$, z, $\mathbf{u}\cdot \mathbf{r}$, and $\Delta r$.  
It also used Nhits$_{p,12}$ and its ratio to Nhits, along with five other variables that made use of the prompt event position and PMT hit information of the delayed event.  The one temporal and four geometric parameters are T$_{\text{rms}}$, N$_{\text{cluster}}$, $\phi_{\text{rms}}$, $\theta_{\text{mean}}$, $\theta_{\text{rms}}$, and are defined in Refs.~\cite{Zhang:2013tua, Zhang:2016tub}.  
The delayed BDT distribution for IBDs is shown as a red histogram in Fig.~\ref{fig:all_plots} (d).  

Cuts on the LR and BDT distributions were determined by maximizing the median discovery significance presented in Ref.~\cite{Galen:2010}, which takes the predicted signal and background counts and background uncertainties as input.  
Backgrounds with delayed events from a neutron capture have fewer discriminating characteristics against IBDs, making the accidental background the most sensitive to the LR or BDT selections and therefore critical in determining their optimal cuts.  
Both LR and BDT methods predicted the accidental background rate using the observed rates of prompt- and delayed-like events.  
The accidental rate was stable across the 190-day lower-background dataset, and, after applying the optimal LR or BDT cuts, was reduced by a factor greater than 10$^{4}$. The expected number of accidental coincidences was 0.7 $\pm$ 0.1 for the LR method after requiring the internal (external) LR $>$ 8.0 (10.4), and 1.4 $\pm$ 0.1 for the BDT method after requiring the internal (external) prompt BDT $>$ 0.1 (0.18) and the delayed BDT $>$ 0.23. 
The IBD efficiencies for these cuts were 45\% and 59\%, and the expected numbers of selected IBDs were $3.5\pm0.7$ and $4.8\pm1.4$.  
These and the expected accidentals are given in Table~\ref{tab:sum_sig_bkg}.  

The estimates of accidental backgrounds were checked by looking at the number of coincidences in a sideband of $\Delta t$ between 500 and 1000~$\mu$s, keeping all other selection criteria unchanged.  
Figures~\ref{fig:all_plots} (a) and (d) show the expected internal LR and delayed BDT distributions for IBDs and accidentals, together with the data from the signal and $\Delta t$ sideband regions. The agreement between the sideband data and accidental expectation is very good for both methods and the total number of accidentals that passed the final cuts in the sideband (0 and 2 for the LR and BDT methods) was consistent with the expectations of 0.7 and 1.4.

\textit{Neutron-capture backgrounds}.  
The \alphan background originates from naturally occurring $\alpha$ decays in which the $\alpha$ particles interact with $^{13}$C in the acrylic of the AV or with $^{18}$O in the AV and water. The $^{13}$C$(\alpha,n)^{16}$O reaction can produce deexcitation $\gamma$'s of up to 6.1~MeV and the $^{18}$O$(\alpha,n)^{21}$Ne reaction produces deexcitation $\gamma$'s of up to 2.8~MeV. 
The $^{13}$C$(\alpha,n)^{16}$O reaction thus produces higher-energy prompt events, but was highly localized to the AV, while the $^{18}$O$(\alpha,n)^{21}$Ne were distributed throughout the detector volume, but produce only lower-energy prompt events. The \alphan reactions were simulated using RAT with cross sections and branching ratios from Refs.~\cite{JENDL/AN-2005, NNDC}. 

The rate of \alphan reactions depends directly on the rate of $\alpha$ decays, which was dominated by $^{210}$Po in both the AV and the water. Predictions of the $^{210}$Po activity were based on radio assays of the AV and water, and they included the rate at which $^{210}$Po leached into the water from the AV based on an {\it ex situ} leaching rate measurement.  
The rate of \alphan around the AV was studied in a FV sideband, using prompt events with positions around the AV in FVs complementary to the IBD FVs (see Table~\ref{tab:cuts}). To increase the statistics of this measurement, all 331 days of live time were used.  
In this sideband, the LR (BDT) method observed 17 (25) events with a background of 2.6 $\pm$ 0.2 (7.1 $\pm$ 0.5) events.  
In both cases the background was dominated by accidentals, which was verified in a $\Delta t$ sideband between 500 and 1000~$\mu$s. The $\Delta t$ distributions were fit with an exponential plus a constant, yielding lifetimes of (276 $\pm$ 96)~$\mu$s and (183 $\pm$ 63)~$\mu$s, respectively.  
These values are consistent with the expectation of 207~$\mu$s.  
The two sideband measurements found consistent rates of \alphan reactions and were averaged to provide a single measured value.  
To our knowledge, this is the first identification of \alphan events in a water Cherenkov detector.

\begin{table}[tbp]\centering
\caption{Expected signal and backgrounds for LR and BDT methods in 190 live days. For BDT, only a 68\% CL upper limit is given for the atmospheric $\nu$ NC background since no events were observed in the sideband. The sum was obtained assuming the limit is the central value, which made the calculated significance conservative.}\label{tab:sum_sig_bkg}
\begin{tabular}{lcc}
\hline
\hline
   &  LR  &  BDT \\
\hline
Reactor IBD\;    & $3.5\pm0.7$  &  $4.8\pm1.4$  \\
Accidentals    & $0.7\pm0.1$  &  $1.4\pm0.1$  \\
($\alpha, n$)  & $0.7\pm0.7$  &  $0.9\pm0.7$  \\
Atm. $\nu$ NC  & \;\;$0.4\pm0.3$\;\;  &  \;$<$0.6\hspace{33pt}  \\ \hline
Sum            & $5.3\pm1.0$  &  $7.7\pm1.7$  \\ \hline
Observation    &  9           &  10           \\ \hline \hline
\end{tabular}
\end{table}

The AV sideband measurement observed a rate of \alphan events that was 4 times higher than predicted, possibly due to a smaller-than-expected rate of leaching from the AV into the water.  In addition to scaling the predicted rate from the AV, a 400\% uncertainty was propagated to the predicted number of \alphan interactions in the water, which was not measured directly. The impact of this large uncertainty on the total \alphan background (including both the AV and water) was relatively small because the water component of the background has low prompt energies and is largely removed by the 2.5-MeV prompt energy criterion (see Table~\ref{tab:cuts}). The statistical uncertainty of the AV sideband measurement dominates the uncertainty on the total \alphan background.  The \alphan background expectations for the LR and BDT methods are given in Table~\ref{tab:sum_sig_bkg}.

Atmospheric neutrinos can interact with oxygen nuclei, ejecting neutrons or protons, and creating excited states of $^{15}$O or $^{15}$N. The ejected nucleons can collide with nuclei and excite them or eject more nucleons.  Neutral-current (NC) interactions are the dominant component of the atmospheric background since charged-current $\nu_{\mu}$ and $\nu_{e}$ interactions produce high-energy leptons, which are easy to identify and reject.  In NC interactions, $^{15}$O* deexcites by emitting high-energy $\gamma$'s~\cite{Ejiri:1993}, creating a prompt signal that is followed by the capture of one or more neutrons by the hydrogen nuclei in the water.

NC atmospheric neutrino interactions produce events with observable energies above those of the IBD prompt events, primarily due to multiple deexcitation $\gamma$'s from more than one nucleus. These interactions also often create multiple neutrons. This background analysis used a sideband with prompt event energies between 2.5 and 25~MeV, in which events in the range of 2.5$-$9.5~MeV (9.5$-$25~MeV) were required to have a neutron multiplicity $\geq$ 2 ($\geq$ 1). As in the \alphan sideband analysis, all 331 days of data were used to increase statistics.  

The LR method observed 2 events in the energy-multiplicity sideband and the BDT method observed 0 events. A GENIE-based simulation~\cite{GENIE:2010} of atmospheric neutrinos was used to estimate the ratio of event rates in the sideband and signal regions, which was used to translate the sideband observations into constraints on the number of events in the signal region. Since no sideband events were observed with the BDT method, an upper limit was given at a 68\% confidence level (CL). Using a ratio between the two regions eliminated uncertainties on the flux and largely on the cross section. 
Systematic uncertainties still arose from uncertainties on the neutron multiplicity and prompt energy distributions.  
These systematics were estimated by comparing the GENIE output to data from  Super-Kamiokande~\cite{Super-Kamiokande:2019hga} and T2K~\cite{T2k:2019NC}, yielding a 5\% uncertainty in the event rate.  
The total uncertainty was dominated by the statistics of the sideband measurement.  
The atmospheric $\nu$ background expectations for the LR and BDT methods are given in Table~\ref{tab:sum_sig_bkg}.

Backgrounds from cosmogenic muons were efficiently avoided by rejecting all data within 20~s after a muon was identified.  This results in no backgrounds from isotopes that undergo $\beta$-$n$ decay, for which the closest candidate is $^{17}$N~\cite{Zhang:2016tub}.  With a lifetime of 4.2~s, the rate of observation was calculated to be $<$ 0.01 per year.

Geoneutrinos from $^{238}$U and $^{232}$Th decays in the Earth were predicted to produce about one quarter of the IBDs produced by reactor $\overline{\nu}_e$'s; however, the number of selected geo-IBDs was negligible because only a small fraction produced a prompt event with an energy above 2.5~MeV.

\textit{Results}.  
After all background sideband analyses were completed, 
the LR and BDT cuts were optimized as described above, and the reactor \nuebar discovery sensitivities were thus estimated to be 1.7$\sigma$ and 2.1$\sigma$.  
Then, the signal region was analyzed in the 190-day dataset, finding 9 coincidences for the LR method and 10 for the BDT method.  
The differences between the observed and expected counts were +3.7 $\pm$ 3.2 (LR) and +2.3 $\pm$ 3.5 (BDT), revealing observations greater than expectations, though not inconsistent.  
The total number of distinct coincidences observed by the two methods was 14, of which 5 were common to both.  
The fraction of all selected IBDs that were selected by both methods was determined with simulations to be 47\%, which was supported by AmBe source data.  
The fraction observed in the data is approximately 5/14 = 36\% (ignoring backgrounds), which is consistent with the expectation.  

The events were directly inspected by fitting an exponential plus a constant to the $\Delta t$ distributions, which yielded fitted lifetimes of (169 $\pm$ 78)~$\mu$s and (207 $\pm$ 82)~$\mu$s for the LR and BDT methods, respectively.  These fits are shown in Figs.~\ref{fig:all_plots} (b) and (e), and are consistent with the expected neutron capture time. 
Figures~\ref{fig:all_plots} (c) and (f) show the observed prompt energy distributions, together with the expectations for the backgrounds.  
The selected events are distributed uniformly across the detector and across calendar time, and the delayed event energy, $\Delta r$, and prompt and delayed $\mathbf{u}\cdot \mathbf{r}$ distributions are also consistent with the simulation~\cite{supp}. 
The 141-day higher-background dataset was also analyzed, revealing 3 and 1 coincidences for the LR and BDT methods, which are consistent with expectations of $2.0\pm0.4$ and $2.3\pm0.5$.  

A discovery significance of reactor \nuebar was calculated just as the sensitivity was, except that the predicted sum of signal and background events was replaced with the observed number.  The resulting significances were 3.0$\sigma$ (LR) and 2.9$\sigma$ (BDT).  
A combined significance was calculated in the same way but using the total number of distinct coincidence events observed by the two methods (14) and estimating the total number of distinct background events and its uncertainty.  
The fraction of all background events that were selected by both methods was assumed to be the same as the fraction of IBDs; i.e., 47\%. 
This assumption is based on the prediction that most backgrounds have a signal-like delayed event; namely, due to a neutron.  The corresponding background uncertainty was conservatively estimated by assuming full correlation ($\rho$ = 1) between the uncertainties of the two methods, finally yielding an expectation of 3.2 $\pm$ 1.0 background events.  
This gives a combined discovery significance of 3.5$\sigma$.  Changes in the assumptions, such as taking accidentals to have no events common to both methods, or assuming a correlation of $\rho$ = 0.5 for the background uncertainties, yielded a significance in the range of 3.2$\sigma$ to 3.7$\sigma$.  

In the absence of oscillations, the expected number of IBDs would approximately double, yielding a signal plus background around 8 and 11 events for the LR and BDT methods, respectively. As such, the current observations cannot distinguish between the oscillation and no-oscillation hypotheses.

\textit{Conclusion}.  
With a detector energy threshold around 1.4~MeV, the SNO+ collaboration has performed the lowest-energy analysis in a large water Cherenkov detector.  
In a search for \nuebar from reactors at least 240~km away, two analytical methods suppressed the accidental background by more than 4 orders of magnitude and made sideband measurements for the three relevant backgrounds.  
With 190 days of data, the two methods obtained consistent evidence for reactor \nuebar and yielded a combined significance of 3.5$\sigma$, producing the first evidence of reactor \nuebar in a Cherenkov detector.

\begin{acknowledgments}
Capital construction funds for the SNO\raisebox{0.5ex}{\tiny\textbf{+}} experiment were provided by the Canada Foundation for Innovation (CFI) and matching partners. 
This research was supported by: 
{\it Canada: }
Natural Sciences and Engineering Research Council, 
the Canadian Institute for Advanced Research (CIFAR), 
Queen's University at Kingston, 
Ontario Ministry of Research, Innovation and Science, 
 Alberta Science and Research Investments Program, 
Federal Economic Development Initiative for Northern Ontario,
Ontario Early Researcher Awards;
{\it U.S.: }
Department of Energy Office of Nuclear Physics, 
National Science Foundation, 
Department of Energy National Nuclear Security Administration through the Nuclear Science and Security Consortium; 
{\it UK: }
Science and Technology Facilities Council (STFC),
the European Union's Seventh Framework Programme under the European Research Council (ERC) grant agreement,
the Marie Curie grant agreement;
{\it Portugal: }
Funda\c{c}\~{a}o para a Ci\^{e}ncia e a Tecnologia (FCT-Portugal);
{\it Germany: }
the Deutsche Forschungsgemeinschaft;
{\it Mexico: }
DGAPA-UNAM and Consejo Nacional de Ciencia y Tecnolog\'{i}a; 
{\it China: }
Discipline Construction Fund of Shandong University. 

We thank the SNO\raisebox{0.5ex}{\tiny\textbf{+}} technical staff for their strong contributions.  We would like to thank SNOLAB and its staff for support through underground space, logistical and technical services. SNOLAB operations are supported by CFI and the Province of Ontario Ministry of Research and Innovation, with underground access provided by Vale at the Creighton mine site.

This research was enabled in part by support provided by WestGRID~\cite{westgrid} and Compute Canada~\cite{computeca} in particular computer systems and support from the University of Alberta~\cite{ualberta} and from Simon Fraser University~\cite{sfu} and by the GridPP Collaboration, in particular computer systems and support from Rutherford Appleton Laboratory~\cite{gridpp, gridpp2}. Additional high-performance computing was provided through the ``Illume'' cluster funded by CFI and Alberta Economic Development and Trade (EDT) and operated by ComputeCanada and the Savio computational cluster resource provided by the Berkeley Research Computing program at the University of California, Berkeley (supported by the UC Berkeley Chancellor, Vice Chancellor for Research, and Chief Information Officer). Additional long-term storage was provided by the Fermilab Scientific Computing Division. Fermilab is managed by Fermi Research Alliance, LLC (FRA) under Contract with the U.S. Department of Energy, Office of Science, Office of High Energy Physics.

For the purposes of open access, the authors have applied a Creative Commons Attribution licence to any Author Accepted Manuscript version arising. Representations of the data relevant to the conclusions drawn here are provided within this paper.
\end{acknowledgments}

\bibliographystyle{apsrev4-2} 
\bibliography{References}

\end{document}